\documentclass[journal]{IEEEtran}
\usepackage[utf8]{inputenc}
\usepackage[T1]{fontenc}

\ifCLASSINFOpdf
\else
   \usepackage[dvips]{graphicx}
\fi

\usepackage{cite}

\usepackage{graphicx}
\usepackage{array}
\usepackage{graphics}
\usepackage{amssymb}
\usepackage{amsmath}
\usepackage{xcolor}

\usepackage[ruled,vlined]{algorithm2e}
\usepackage{algorithmic}
\usepackage{mathtools}

\DeclareMathOperator*{\argmin}{arg\,min\!}
\newcommand{\pinv}{\dag}
\newcommand{\ma}[1]{\boldsymbol{#1}}
\newcommand{\compl}{\mathbb{C}}
\newcommand{\real}{\mathbb{R}}
\newcommand{\diagof}[1]{{\rm diag\!}\left\{ #1 \right\}}
\newcommand{\opvec}[1]{{\rm vec}\left\{ #1 \right\}}
\newcommand{\vecd}[1]{{\rm vecd}\left\{ #1 \right\}}
\newcommand{\ten}[1]{\mathcal #1}
\newcommand{\fronorm}[1]{\left\|#1\right\|_{\rm F}}
\newcommand{\fronormbig}[1]{\Bigl|\Bigl|#1\Bigr|\Bigr|_{\rm F}}
\newcommand{\krp}{\diamond}
\newcommand{\kron}{\otimes}
\newcommand{\trans}{{{\rm T}}}
\newcommand{\herm}{{{\rm H}}}
\newcommand{\unf}[2]{\left[ \ten{#1} \right]_{(#2)}}
\newcommand{\mat}[1]{\ensuremath{{\mathbf{#1}}}}
\newcommand{\vet}[1]{\mathit{\boldsymbol{#1}}}
\newcommand{\lowint}[1]{\left\lceil{#1}\right\rceil}

\renewcommand\baselinestretch{.96}

\begin{document}
\bstctlcite{IEEEexample:BSTcontrol}
\title{Reducing Complexity of Data-Aided Channel Estimation in RIS-Assisted Communications
\thanks{Amarilton L. Magalhães and André L. F. de Almeida are with Department of Teleinformatics Engineering, Federal University of Ceará, Brazil. Gilderlan T. de Araújo is with Federal Institute of Ceará, Brazil. E-mails: \{amarilton,andre,gilderlan\}@gtel.ufc.br. The authors thank the partial support of National Institute of Science and Technology (INCT-Signals) under the grant INCT-25255-82587.32.41/64, sponsored by FUNCAP/CNPq/CAPES, Brazil.}
}
\author{Amarilton L. Magalhães, André L. F. de Almeida, and Gilderlan T. de Araújo\\[4ex]}

\maketitle

\begin{abstract}
We consider the data-aided channel estimation (CE) problem in a reconfigurable intelligent surface (RIS)-assisted wireless communication system, where the channel and information symbols are estimated jointly during the CE phase, differently from pure pilot-aided methods. We propose a two-stage semi-blind receiver that jointly estimates the combined channel and the data symbols, followed by channel decoupling. To this end, we derive a new modeling framework whose first stage recasts the received signal to allow for the joint estimation of the combined channel and transmitted symbols. In the second stage, channel decoupling is easily achieved \emph{via} Khatri-Rao factorization, yielding a refined channel estimate. Our solution yields accurate estimates of the cascaded channel at lower computational complexity. Simulation results reveal a similar performance of the proposed method to that of the competitor while providing a substantially reduced computational cost.
\end{abstract}
\begin{IEEEkeywords}
Reconfigurable intelligent surfaces, data-aided channel estimation, tensor modeling.
\end{IEEEkeywords}
\IEEEpeerreviewmaketitle

\section{Introduction}
\IEEEPARstart{R}{}econfigurable intelligent surface (RIS) is a panel comprised of a large array of (semi)passive reflecting elements capable of shaping the wireless propagation environment by independently modifying the phase of the impinging electromagnetic waves \cite{basar2019wireless,you2025next}. This feature enables the RIS to expand coverage areas while providing energy-efficient communication, being a strong candidate for future wireless network deployments, such as the sixth generation (6G) \cite{katwe2024overview}. However, obtaining precise channel state information at an affordable computational cost is crucial and challenging in properly optimizing the RIS and the transmit/receive beamforming at the base station (BS) and user terminal (UT). To deal with this task, channel estimation (CE) methods have been studied in the literature (see \cite{yan2025power, sun2025power, swindlehurst2022channel, zhou2022channel, dearaujo2023semiblind,dearaujo2021channel} and references therein).

Tensor approaches have been employed in CE for RIS-assisted wireless systems \cite{dearaujo2021channel,xu2022sparse, dearaujo2023semiblind, benicio2024tensor_wcl}. Such works harness the multi-linear structure of the signals and showcase unique properties of tensor decompositions in enabling receivers with fewer restrictions and balanced performance/complexities. Recently, the work \cite{dearaujo2023semiblind} proposed an iterative data-aided semi-blind receiver with integrated joint symbol and CE estimation features for a MIMO wireless communication system assisted by a RIS. Therein, the authors exploited the tensor structure of the received signals according to PARATUCK tensor decomposition \cite{FavierAlmeida2014, favier2012tensor}. In contrast to only pilot-aided techniques, such as in \cite{dearaujo2021channel}, the data-aided method proposed by \cite{dearaujo2023semiblind} can efficiently solve the CE problem while detecting transmitted information symbols. This approach reduces the symbol decoding delay and can improve the data rate since such symbols are estimated in advance, still at the CE stage. However, \cite{dearaujo2023semiblind} alternately estimates three matrices (symbol and UT-RIS/RIS-BS channel matrices) and refines such estimates at each iteration, implying a high computational cost.

This letter shows that the data-aided CE problem for RIS-assisted communications \cite{dearaujo2023semiblind} can efficiently be achieved with a semi-blind receiver with significantly lower computational complexity. We propose a two-stage approach. In the first stage, by exploiting the received signal structure, we propose an algebraic formulation in which the symbol matrix is alternately estimated as a function of the combined channel instead of the individual channel matrices \cite{dearaujo2023semiblind} and vice-versa, according to an iterative linear estimation process. In the second stage, channel decoupling is easily achieved \emph{via} Khatri-Rao factorization, yielding refined channel estimates. To our knowledge, the proposed approach has not yet been explored in the channel estimation literature for RIS-assisted systems. Simulation results show that the proposed method performs similarly to the competitor with a substantially reduced cost.

\

\vspace{-1.5ex}\noindent\textit{Notations and properties}:
We use $a$, $\vet{a}$, $\mat{A}$, and $\ten{A}$ for scalars, vectors, matrices, and tensors, respectively. $\mat{A}^\trans$, $\mat{A}^\herm$, and $\mat{A}^\pinv$ stem for transpose, Hermitian and Moore-Penrose pseudo-inverse of $\mat{A}$. The Frobenius norm is represented by $\fronorm{\cdot}$, and the Kronecker and Khatri-Rao matrix products are denoted by $\kron$ and $\krp$, respectively. Given $\mat{A} \in \compl^{I \times J}$, the vectorization operator $\opvec{\mat{A}}$ yields $\vet{a} \in \compl^{JI \times 1}$, and the reverse operation $\mathrm{unvec}_{I \times J}(\vet{a})$ brings $\mat{A}$ back. In addition, $\diagof{\vet{a}}$ forms a diagonal matrix from $\vet{a}$. If $\mat{B}$ is a diagonal matrix, $\mathrm{vecd}\{\mat{B}\}$ returns a vector out of the diagonal of $\mat{B}$. In this paper, we make use of the following identities:
\begin{gather}
	\label{prop:vec} \opvec{\mat{ABC}} = (\mat{C}^\trans \kron \mat{A})\opvec{\mat{B}};\\
    \label{prop:vecd} \opvec{\mat{ABC}} = \left(\mat{C}^\trans \krp \mat{A}\right)\vecd{\mat{B}}, \quad \mathrm{\mbox{for }} \mat{B} \mathrm{\mbox{ diagonal}};\\
    \label{prop:kronkron} \mat{AB} \kron \mat{CD} = (\mat{A} \kron \mat{C})(\mat{B} \kron \mat{D}).
\end{gather}

Stated $\mat{A} \in \compl^{P \times Q}$ and $\mat{B} \in \compl^{I \times J}$, we adopt the following equivalence for the vectorization of $\mat{A} \kron \mat{B} \in \compl^{PI \times QJ}$ \cite{teseflorian}:
\begin{equation}\label{prop:vecmatab}
	\opvec{\mat{A} \kron \mat{B}} = (\mat{I}_Q \kron \bar{\mat{B}})\opvec{\mat{A}} \in \compl^{QJPI \times 1},
\end{equation}
where $\bar{\mat{B}} \doteq [\mat{I}_P \kron \vet{b}_1^\trans,\cdots,\mat{I}_P \kron \vet{b}_J^\trans]^\trans \in \compl^{JPI \times P}$. In addition, we use the definition of permuted-column Kronecker product by solely reordering the columns of $\mat{A} \kron \mat{B}$, given by \cite{teseflorian}
\begin{equation}\label{pckron}
    \mat{A} \boxtimes \mat{B} = [\mat{A} \kron \vet{b}_1,\cdots,\mat{A} \kron \vet{b}_J] \in \compl^{PI \times JQ}.
\end{equation}

\section{System and signal models}\label{sec:sys}
We adopt the uplink RIS-assisted MIMO communication model consisting of a UT with $L$ antennas communicating with a BS equipped with $M$ antennas through a RIS with $N$ elements. Due to blockage in the direct link, non-line-of-sight communication is assumed. The UT encodes each data stream by following a Khatri-Rao space-time coding scheme \cite{Sidropoulos_2002_KRST, dearaujo2023semiblind} before transmission, which follows a two-block time structure. The first span $K$ sub-frames of $T$ symbols each and is dedicated to data-aided CE, while the second is for pure data decoding using the estimated channels. We consider a quasi-static flat-fading channel with a coherence time of $KT$ symbol periods. In the first block, the UT transmits data symbols that may change within the sub-frame, while the coding and the RIS phase shifts remain fixed within sub-frames but may vary across them. This paper focuses solely on the first block. The received signal at BS over $T$ symbols is \cite{dearaujo2023semiblind}
\begin{equation}\label{matyk}
    \mat{Y}_k = \mat{H}\diagof{\vet{\psi}_k}\mat{G}\diagof{\vet{\lambda}_k}\mat{X} + \mat{V}_k  \in \compl^{M \times T},
\end{equation}
where $\mat{H} \in \compl^{M \times N}$ is the RIS-BS channel matrix, $\vet{\psi}_k \in \compl^{N \times 1}$ is phase-shift vector of the $k$-th sub-frame, $\mat{G} \in \compl^{N \times L}$ is the UT-RIS channel matrix, $\vet{\lambda}_k \in \compl^{L \times 1}$ is the coding vector related to the $k$-th sub-frame, $\mat{X} \in \compl^{L \times T}$ is the symbol matrix and $\mat{V}_k  \in \compl^{M \times T}$ is the corresponding additive noise term. Since our method does not rely on any a priori channel knowledge, optimization of RIS phase-shifts and coding as a function of the channel state information cannot be carried out. Nevertheless, to optimize the CE performance, we have designed the coding and the phase-shifts based on a discrete Fourier transform (DFT) matrix, following the work \cite{jensen2020optimal}. 

Without the direct link, the work \cite{dearaujo2023semiblind} proposed a trilinear alternating least-squares (TALS) semi-blind receiver to solve the data-aided CE problem, which consists of iteratively estimating the UT-RIS and RIS-BS channels $\mat{H}$ and $\mat{G}$ jointly with the data symbol matrix $\mat{X}$ by exploiting the structure of the data tensor $\mathcal{Y} \in \mathbb{C}^{M \times T \times K}$ built from the set of received signal matrices $\{\mat{Y}_k\}$, $k=1, \ldots, K$, given in $(\ref{matyk})$. The works \cite{dearaujo2021channel} and \cite{dearaujo2023semiblind} show that estimating the individual channels before reconstructing the combined one yields significant performance gains over classical LS methods. In contrast, our proposed solution first estimates the combined channel using actual data symbols and defers the channel decoupling to a second stage, ensuring similar performance while significantly reducing the computational cost, as detailed in the next section.

\section{Two-stage semi-blind (TSB) receiver}
The proposed data-aided semi-blind receiver has two stages. In the first stage, we aim to jointly estimate the combined channel (encapsulating the UT-RIS and RIS-BS channels) and the symbol matrix. Let us consider modifying the problem 
\begin{equation}\label{mintals}
     \min_{\mat{H},\mat{G},\mat{X}} \quad \sum\limits_{k=1}^K\fronormbig{\mat{Y}_k - \mat{H}\diagof{\vet{\psi}_k}\mat{G}\diagof{\vet{\lambda}_k}\mat{X}}^2,
\end{equation}
tackled in \cite{dearaujo2023semiblind}. We rewrite \eqref{matyk} by applying the $\mathrm{vec}\{\cdot\}$ operator along with property \eqref{prop:vecd} to yield
\begin{equation}\label{ykthetapsik}
    \vet{y}_k = \bigl(\mat{X}^\trans\diagof{\vet{\lambda}_k} \kron \mat{I}_M\bigr)\ma{\Theta}\vet{\psi}_k + \vet{v}_k,
\end{equation}
where $\vet{y}_k \doteq \opvec{\mat{Y}_k} \in \compl^{TM \times 1}$, $\vet{v}_k \doteq \opvec{\mat{V}_k} \in \compl^{TM \times 1}$, while $\ma{\Theta} \doteq \mat{G}^\trans \krp \mat{H} \in \compl^{LM \times N}$ is the Khatri-Rao structured combined channel matrix. Then, we propose recast the optimization problem \eqref{mintals} to 
\begin{equation}\label{minbals}
     \min_{\ma{\Theta},\mat{X}} \quad \sum\limits_{k=1}^K\fronormbig{\vet{y}_k - \bigl(\mat{X}^\trans\diagof{\vet{\lambda}_k} \kron \mat{I}_M\bigr)\ma{\Theta}\vet{\psi}_k}^2.
\end{equation}

By manipulating the corresponding matrix unfoldings, we can obtain an interesting algebraic formulation in which the symbol matrix is alternately estimated as a function of the combined channel and vice versa, according to an iterative linear estimation process. Unlike the semi-blind TALS receiver \cite{dearaujo2023semiblind} that seeks to estimate three quantities ($\hat{\mat{H}}$, $\hat{\mat{G}}$, and $\hat{\mat{X}}$), implying a computationally expensive iterative process and slow convergence\cite{comon2009tensor}, the proposed problem \eqref{minbals} only involves estimating two quantities, namely $\hat{\ma{\Theta}}$ and $\hat{\mat{X}}$, which can be accomplished fastly, while exhibiting reduced computational cost. Explicit expressions for estimating the combined channel and the symbol matrix can be obtained after some algebraic manipulations done in Subsections \ref{subsec:thetaest} and \ref{subsec:xest}. Upon completion of this first stage, the second one consists of decoupling the estimates of the individual channel matrices. For this purpose, we consider the optimization problem
\begin{equation}\label{minkhatri}
    \min_{\mat{H},\mat{G}} \fronorm{\hat{\ma{\Theta}} - \mat{G}^\trans \krp \mat{H}}^2.
\end{equation}
To tackle \eqref{minkhatri}, Khatri-Rao factorization \cite{dearaujo2021channel,Kibangou2009} can be easily applied to $\hat{\ma{\Theta}}$ to extract separate estimates of $\mat{H}$ and $\mat{G}$. We describe this process in Section \ref{subsec:krfstage}.

\subsection{Estimation of the combined channel}\label{subsec:thetaest}
Applying $\mathrm{vec}(\cdot)$ to \eqref{ykthetapsik} and using the property \eqref{prop:vec}, we obtain $\vet{y}_k = \bigl(\vet{\psi}_k^\trans \kron \mat{X}^\trans\diagof{\vet{\lambda}_k} \kron \mat{I}_M\bigr)\vet{\theta} + \vet{v}_k$, where $\vet{\theta} \doteq \mathrm{vec}\{\ma{\Theta}\} \in \compl^{NLM \times 1}$. Now, from property \eqref{prop:kronkron}, we have $\vet{y}_k = \bigl[\mat{X}^\trans\bigl(\vet{\psi}_k^\trans \kron \diagof{\vet{\lambda}_k}\bigr) \kron \mat{I}_M\bigr]\vet{\theta} + \vet{v}_k$. By stacking the set of signal vectors $\{\vet{y}_k\}$, for $k=1,\cdots,K$, we obtain $\mathrm{vec}\bigl\{\unf{Y}{3}^\trans\bigr\} \doteq \bigl[\vet{y}_1^\trans,\cdots,\vet{y}_K^\trans\bigr]^\trans \in \compl^{KTM \times 1}$, which can be expressed in compact form as
\begin{equation}\label{y3t}
    \opvec{\unf{Y}{3}^\trans} = \bigl(\mat{F}(\mat{X}) \kron \mat{I}_M\bigr)\vet{\theta} + \vet{v},
\end{equation}
where $\unf{Y}{3}$ denotes the 3-mode unfolding of the received signal tensor, $\mat{F}(\mat{X})\doteq (\mat{I}_K \kron \mat{X}^\trans)\mat{Z} \in \compl^{KT \times NL}$, with $\mat{Z} \!\doteq\! \bigl[\vet{\psi}_1 \!\kron \diagof{\vet{\lambda}_1}\!,\!\cdots\!,\!\vet{\psi}_K \!\kron \diagof{\vet{\lambda}_K}\bigr]^\trans \!\!\in\! \compl^{KL \times NL}$, and $\vet{v} \in \compl^{KTM \times 1}$ is the corresponding noise term. Hence, an LS estimate of $\vet{\theta}$ can be obtained by solving the problem
\begin{equation}
	\hat{\vet{\theta}} \!=\! \argmin_{\vet{\theta}} \fronorm{\mathrm{vec}\bigl\{\unf{Y}{3}^\trans\bigr\} \!-\! \bigl(\mat{F}(\mat{X}) \kron \mat{I}_M\bigr)\vet{\theta}}^2,
\end{equation}
whose analytical solution is given by
\begin{equation}\label{thetaest}
	\hat{\ma{\Theta}} \!=\! \mathrm{unvec}_{LM \times N}\Bigl\{\!\Bigl(\!\mat{F}(\mat{X})^\pinv \!\kron \mat{I}_M\!\Bigr)\mathrm{vec}\bigl\{\unf{Y}{3}^\trans\bigr\}\!\Bigr\}.
\end{equation}

\subsection{Symbol estimation}\label{subsec:xest}
Deriving an estimate of $\mat{X}$ as a function of $\ma{\Theta}$ is an involved task. However, we demonstrate this by employing useful identities \cite{teseflorian}. To begin, consider the transpose of \eqref{ykthetapsik}, i.e., $\vet{y}_k^\trans = \vet{\psi}_k^\trans\ma{\Theta}^\trans\bigl(\diagof{\vet{\lambda}_k}\mat{X} \kron \mat{I}_M\bigr) + \vet{v}_k^\trans \in \compl^{1 \times TM}$. By applying $\opvec{\cdot}$ along with property \eqref{prop:vec}, we get
\begin{equation}
	\vet{y}_k = \bigl(\mat{I}_{TM} \kron \vet{\psi}_k^\trans\ma{\Theta}^\trans\bigr)\opvec{\diagof{\vet{\lambda}_k}\mat{X} \kron \mat{I}_M} + \vet{v}_k.
\end{equation}
The next step is to apply the identity \eqref{prop:vecmatab} to $\opvec{\diagof{\vet{\lambda}_k}\mat{X} \kron \mat{I}_M} \in \compl^{TMLM \times 1}$ that yields
\begin{equation}
	\vet{y}_k = \bigl(\mat{I}_{TM} \kron \vet{\psi}_k^\trans\ma{\Theta}^\trans\bigr)\bigl(\mat{I}_T \kron \mat{J}\bigr)\mathrm{vec}\bigl\{\diagof{\vet{\lambda}_k}\mat{X}\bigr\} + \vet{v}_k,
\end{equation}
where $\mat{J} \doteq \bigl[\mat{I}_L \kron \vet{e}_1^\trans,\cdots,\mat{I}_L \kron \vet{e}_M^\trans\bigr]^\trans \in \real^{MLM \times L}$, with $\vet{e}_m \!\in\! \real^{M \times 1}$ ($m=1,\cdots,M$) being the $m$-th column of $\mat{I}_M$. Defining $\vet{x} \doteq \opvec{\mat{X}} \in \compl^{TL \times 1}$ and utilizing \eqref{prop:vec}, we get
\begin{equation*}\label{ykititit}
	\vet{y}_k = \Bigl[\mat{I}_T \kron \bigl(\mat{I}_M \kron \vet{\psi}_k^\trans\ma{\Theta}^\trans\bigr)\Bigr]\bigl(\mat{I}_T \kron \mat{J}\bigr)\bigl(\mat{I}_T \kron \diagof{\vet{\lambda}_k}\bigr)\vet{x} + \vet{v}_k,
\end{equation*}
which is simpliflied by employing the property \eqref{prop:kronkron}, as follows
\begin{equation}
	\vet{y}_k = \bigl[\mat{I}_T \kron \bigl(\mat{I}_M \kron \vet{\psi}_k^\trans\ma{\Theta}^\trans\bigr)\mat{J} \diagof{\vet{\lambda}_k}\bigr]\vet{x} + \vet{v}_k.
\end{equation}
Using \eqref{pckron}, we define the commutation matrices $\mat{P}^{LM}_{M} \!=\! \mat{I}_{LM} \boxtimes \mat{I}_M \!\in\! \compl^{LMM \times MLM}$ and $\mat{P}^M_{LM} \!=\! \mat{I}_M \boxtimes \mat{I}_{LM} \!\in\! \compl^{MLM \times LMM}$, in which $\mat{P}^{LM}_M\mat{P}^M_{LM} \!=\!  \mat{I}_{LMM}$. Using $\mat{P}^{LM}_M$, we obtain the following useful equivalences $\mat{I}_M \kron \ma{\Psi}_{k\cdot}\ma{\Theta}^\trans \!=\! \bigl(\ma{\Psi}_{k\cdot}\ma{\Theta}^\trans \kron \mat{I}_M\bigr)\mat{P}^{LM}_M$ and $[\mat{I}_L \kron \vet{e}_1^\trans,\!\cdots\!,\mat{I}_L \kron \vet{e}_M^\trans]^\trans \!=\! \mat{P}^M_{LM}(\mat{I}_L \kron \vet{i}_M)$, where $\vet{i}_M \!\doteq\! [\vet{e}_1^\trans,\cdots,\vet{e}_M^\trans]^\trans \!=\! \opvec{\mat{I}_M} \!\in\! \real^{MM \times 1}$. Thus,
\begin{equation}\label{eq:y_beforefinal}
	\vet{y}_k = \Bigl[\mat{I}_T \kron \bigl(\vet{\psi}_k^\trans\ma{\Theta}^\trans \kron \mat{I}_M\bigr)(\mat{I}_L \kron \vet{i}_M)\diagof{\vet{\lambda}_k}\Bigr]\vet{x} + \vet{v}_k.
\end{equation}
It is easy to see that $(\mat{I}_L \kron \vet{i}_M)\diagof{\vet{\lambda}_k} = \diagof{\vet{\lambda}_k} \kron \vet{i}_M$. Therefore, \eqref{eq:y_beforefinal} can be compactly expressed as $\vet{y}_k = \bigl(\mat{I}_T \kron \mat{E}_k\bigr)\vet{x} + \vet{v}_k$, where $\mat{E}_k \doteq (\vet{\psi}_k^\trans\ma{\Theta}^\trans \kron \mat{I}_M)(\diagof{\vet{\lambda}_k} \kron \vet{i}_M) \in \compl^{M \times L}$. Using the $\mathrm{unvec}(\cdot)$ operator and the reverse action of \eqref{prop:vec}, we get $\mat{Y}_k = \mathrm{unvec}_{M \times T}\{\vet{y}_k\} = \mat{E}_k\mat{X} + \mat{V}_k$. Stacking row-wise the matrix slices $\mat{Y}_k$ for $k=1,\cdots,K$ yields
\begin{equation}
	\unf{Y}{2}^\trans \doteq \bigl[\mat{Y}_1^\trans,\cdots,\mat{Y}_K^\trans\bigr]^\trans = \mat{E}(\ma{\Theta})\mat{X} + \mat{V} \in \compl^{KM \times T},
\end{equation}
where $\unf{Y}{2}$ denotes the 2-mode unfolding of $\ten{Y}$, $\mat{V} \in \compl^{KM \times T}$ is the associated noise part, and $\mat{E}(\ma{\Theta}) \!\doteq\! \bigl[\mat{E}_1^\trans,\!\cdots\!,\mat{E}_K^\trans\bigr]\!^\trans \in \compl^{KM \times L}$. An estimate of $\mat{X}$ can be obtained by solving the following problem
\begin{equation}
	\hat{\mat{X}} = \argmin_{\mat{X}}\fronorm{\unf{Y}{2}^\trans - \mat{E}(\ma{\Theta})\mat{X}}^2,
\end{equation}
whose analytical solution is
\begin{equation}\label{xest}
	\hat{\mat{X}} = \mat{E}(\ma{\Theta})^\pinv\unf{Y}{2}^\trans.
\end{equation}

Considering $\mat{E}(\mat{G,H}) \!\doteq\! \bigl[\diagof{\vet{\lambda}_1}\mat{G}^\trans\diagof{\vet{\psi}_1}\mat{H}^\trans,$ $\cdots\!,\diagof{\vet{\lambda}_K\!}\mat{G}^\trans\diagof{\vet{\psi}_K\!}\mat{H}^\trans\bigr]^\trans\! \in \compl^{KM \times L}$, defined in \cite{dearaujo2023semiblind} (see Eq. (29) therein), observe that $\mat{E}(\ma{\Theta}) \equiv \mat{E}(\mat{G,H})$. The key difference is that $\mat{E}(\mat{G,H})$ contains the individual channel matrices ($\mat{G}$ and $\mat{H}$), while $\mat{E}(\ma{\Theta})$ is directly expressed as a function of the combined channel $\ma{\Theta}$. 

The proposed two-stage semi-blind (TSB) receiver leverages \eqref{thetaest} and \eqref{xest} to iteratively estimate the combined channel and symbol matrix using a simple alternating linear estimation scheme known as bilinear alternating least squares (BALS). This procedure corresponds to the first stage of Algorithm \ref{alg:balskrf}.

\subsection{Khatri-Rao factorization stage}\label{subsec:krfstage}
The $n$-th column of $\ma{\Theta}$, defined as $\vet{\theta}_n \in \compl^{LM \times 1}$, is given by $\vet{\theta}_n = \vet{g}_n \kron \vet{h}_n = \opvec{\vet{h}_n\vet{g}_n^\trans}$, where $\vet{g}_n \in \compl^{L \times 1}$ and $\vet{h}_n \in \compl^{M \times 1}$ are the $n$-th column of $\mat{G}^\trans$ and $\mat{H}$, respectively \cite{Kibangou2009}. The estimates of the individual channels are found as
\begin{equation}\label{ghestkrp}
	[\hat{\mat{G}},\hat{\mat{H}}] = \argmin_{\vet{g}_n,\vet{h}_n}\sum\limits_{n=1}^N\fronorm{\ma{\Omega}_n - \vet{h}_n\vet{g}_n^\trans}^2,
\end{equation}
where $\ma{\Omega}_n = \mathrm{unvec}_{M \times L}\{\vet{\theta}_n\!\}$ is a rank-one matrix. The estimates of $\hat{\mat{G}}$ and $\hat{\mat{H}}$ are found through $N$ independent rank-one approximation problems, wherein $\hat{\vet{h}}_n$ and $\hat{\vet{g}}_n$ are obtained, respectively, from the dominant left and right singular vectors of $\ma{\Omega}_n$ after a truncated singular value decomposition (SVD). After the KRF, $\hat{\mat{G}}$ and $\hat{\mat{H}}$ are used to reconstruct $\hat{\ma{\Theta}}$. This procedure corresponds to the KRF stage in Algorithm \ref{alg:balskrf}.

\subsection{Computational complexity} \label{subsec:complexity}
The computational complexity of the TSB receiver is dominated by the generalized left-inverses of steps 3 and 4 of Algorithm \ref{alg:balskrf}, which corresponds to the BALS stage, followed by the complexity associated with the computation of $N$ rank-one SVDs, corresponding to the KRF stage. Considering a cost $\mathcal{O}(IJ^2)$ to calculate the pseudo-inverse of a rank-$J$ matrix $\mat{A} \in \compl^{I \times J}$, the complexity of \eqref{thetaest} and \eqref{xest} are, respectively, $\mathcal{O}(KTL^2N^2)$ and $\mathcal{O}(KML^2)$. This means that each iteration of the BALS stage requires $\mathcal{O}(KL^2(TN^2 + M))$. The complexity of the KRF stage involves computing the SVD of each rank-1 matrix $\ma{\Omega}_n$, which implies $\mathcal{O}(ML)$. By considering $N$ parallel SVDs, the total cost of the KRF is $\mathcal{O}(NML)$.

The complexity of the TSB can be significantly reduced by designing $\ma{\Lambda}\!=\![\vet{\lambda}_1,\!\cdots\!,\vet{\lambda}_K]^\trans \!\in\! \compl^{K \times L}$ and $\ma{\Psi}\!=\![\vet{\psi}_1,\!\cdots\!,\vet{\psi}_K]^\trans \!\in\! \compl^{K \times N}$ as semi-unitary matrices. Following \cite{dearaujo2023semiblind}, starting from a truncated DFT matrix $\ma{\Omega} \!\in\! \compl^{LN \!\times\! K}$ ($K \!\geq\! LN$), we extract $\ma{\Lambda}$ and $\ma{\Psi}$ by invoking an exact Khatri-Rao factorization of a DFT matrix (see \cite{sokal2020semi} for details). This is accomplished by doing $\vet{\lambda}_{k+1}^\trans \!=\! [1,\omega^k,\!\cdots\!,\omega^{kN(L-1)}]$ and $\vet{\psi}_{k+1}^\trans \!=\! [1,\omega^k,\!\cdots\!,\omega^{k(N-1)}]$, where $\omega \!=\! e^{-j2\pi/\!K}$. This way, by avoiding matrix inverses, \eqref{thetaest} and \eqref{xest} can be replaced by the simplified expressions
\begin{equation*}
	\hat{\ma{\Theta}} \!=\! (1/K)\mathrm{unvec}_{LM \times N}\Bigl\{\!\Bigl(\diagof{\vet{\zeta}}\mat{F}(\mat{X})^\herm \kron \mat{I}_M\!\Bigr)\mathrm{vec}\bigl\{\unf{Y}{3}^\trans\bigr\}\!\Bigr\},
\end{equation*}
\begin{equation*}
    \hat{\mat{X}} = (1/K)\diagof{\vet{\xi}}\mat{E}(\ma{\Theta})^\herm\unf{Y}{2}^\trans,
\end{equation*}
where $\vet{\zeta} \!=\! \vet{1}_N \!\kron\! \left[1/\|\vet{x}_1\|^2,\!\cdots\!,1/\|\vet{x}_L\|^2\right] \!\in\! \compl^{NL \times 1}$, with $\vet{x}_\ell \!\in\! \compl^{T \times 1}$ denoting the $\ell$-th column of $\mat{X}^\trans$, and $\vet{\xi} \!=\! \left[1/\|\bar{\vet{\theta}}_1\|^2,\!\cdots\!,1/\|\bar{\vet{\theta}}_L\|^2\right] \!\in\! \compl^{L \times 1}$, with $\vet{\theta}_\ell \in \compl^{NM \times 1}$ denoting the $\ell$-th column of the rerranged matrix $\bar{\ma{\Theta}}$, obtained by using the mapping operator $\phi \!:\! \compl^{LM \times N} \!\rightarrow\! \compl^{L \times NM}$ such that $\bar{\ma{\Theta}} \!=\! \phi(\ma{\Theta}) \!\in\! \compl^{L \times NM}$ satisfies $[\bar{\ma{\Theta}}]_{\ell,(n-1)M+m} = [\ma{\Theta}]_{(\ell - 1)M+m,n}$ for $\ell \!=\! 1,\!\cdots\!,L$, $m \!=\! 1,\!\cdots\!,M$ and $n \!=\! 1,\!\cdots\!,N$. Note that this cost reduction requires transmitting more sub-frames.
\begin{algorithm}[t]
	\small
	\caption{Two-stage Semi-Blind (TSB) Receiver}
	\label{alg:balskrf}
    \vspace{1ex}
    \hspace{-3ex} \textit{Stage 1: Bilinear Alternating Least-Squares (BALS)}\\
    \begin{algorithmic}
		\small{
			\STATE \hspace{-4ex} 1. Set $i=0$ and initialize $\hat{\mat{X}}_{(i=0)}$ randomly;\\
			\STATE \hspace{-4ex} 2. $i = i + 1$;\\
			\STATE \hspace{-4ex} 3. Get $\hat{\ma{\Theta}}_{(i)} \!=\! \mathrm{unvec}_{LM \times N}\Bigl\{\mat{F}(\mat{X}_{(i-1)})^\pinv\mathrm{vec}\bigl\{\unf{Y}{3}^\trans\bigr\}\Bigr\}$;\\
			\STATE \hspace{-4ex} 4. Get $\hat{\mat{X}}_{(i)} = \mat{E}(\ma{\Theta}_{(i)})^\pinv\unf{Y}{2}^\trans$;\\
			\STATE \hspace{-4ex} 5. Repeat steps 2-5 until convergence;\\
			\STATE \hspace{-4ex} 6. Remove scaling ambiguities.
		}
	\end{algorithmic}
    \hspace{-3ex} \textit{Stage 2: Khatri-Rao Factorization (KRF)}\\
    \begin{algorithmic}
		\small{
			\STATE \hspace{-4ex} 1. \textbf{for} $n=1,\cdots,N$\\
            \STATE \hspace{-4ex} \quad\quad $\ma{\Omega}_n = \mathrm{unvec}_{M \times L}\{\vet{\theta}_n\}$\\
            \STATE \hspace{-4ex} \quad\quad $[\vet{u}_1,\sigma_1,\vet{v}_1] \longleftarrow$ truncated-SVD$(\ma{\Omega}_n)$\\
			\STATE \hspace{-4ex} \quad\quad $\hat{\vet{g}}_n = \sqrt{\sigma_1}\vet{v}_1^\ast$,~ $\hat{\vet{h}}_n = \sqrt{\sigma_1}\vet{u}_1$\\
			\STATE \hspace{-4ex} \quad \textbf{end};\\
			\STATE \hspace{-4ex} 2. $\hat{\mat{G}} \longleftarrow \left[\hat{\vet{g}}_1,\cdots,\hat{\vet{g}}_N\right]^\trans$, ~$\hat{\mat{H}} \longleftarrow [\hat{\vet{h}}_1,\cdots,\hat{\vet{h}}_N]$;\\
			\STATE \hspace{-4ex} 3. Reconstruct the combined channel: $\hat{\ma{\Theta}} = \hat{\mat{G}}^\trans \krp \hat{\mat{H}}$.
		}
	\end{algorithmic}
\end{algorithm}

\section{Identifiability}
\subsection{Uniqueness}
Uniqueness of the LS estimates of $\ma{\Theta}$ and $\mat{X}$ according to \eqref{thetaest} and \eqref{xest}, respectively, requires that $\mat{F}(\mat{X})$ and $\mat{E}(\ma{\Theta})$ have full column-rank to be left-invertible. To stand this, the necessary conditions $KT \geq NL$ and $KM \geq L$ must be satisfied, respectively. Regarding the uniqueness of $\hat{\ma{\Theta}}$, consider the known properties $\mathrm{rank}(\mat{AB}) \leq \mathrm{min}\{\mathrm{rank}(\mat{A}),\mathrm{rank}(\mat{B})\}$ and $\mathrm{rank}(\mat{A} \kron \mat{B}) = \mathrm{rank}(\mat{A})\mathrm{rank}(\mat{B})$. For simplicity, we assume that $\mat{Z}$ is designed to have full column rank (which implies $K \geq N$). Then, $\mathrm{rank}\bigl(\mat{F}(\mat{X})\bigr) \leq \mathrm{min}\{K\mathrm{rank}(\mat{X}),NL\}$. Since $K \geq N$, $\mat{X}$ must have full row-rank to $\mat{F}(\mat{X})$ be full column-rank, which requires $T \geq L$. Otherwise, if $\mat{X}$ is full column-rank, implying $T < L$, the UT must compensate for this by transmitting more sub-frames to ensure $KT \geq NL$.

Let us now consider the uniqueness of $\hat{\mat{X}}$, which depends on the properties of $\ma{\Theta}$. If $\mat{G}$ and/or $\mat{H}$ have full-rank, which corresponds to assuming scattering-rich propagation (e.g. Rayleigh fading) for at least one link, $\ma{\Theta}$ will always have full-rank due to its Khatri-Rao structure. This comes from the fundamental result on the rank of a Khatri-Rao product \cite{sidiropoulos2000uniqueness}, which in our case states that $\mathrm{rank}\bigl(\ma{\Theta}\bigr) \geq \mathrm{rank}\bigl(\mat{G}^\trans \diamond \mat{H}\bigr)\geq \textrm{min}(\mathrm{rank}(\mat{G})+\mathrm{rank}(\mat{H})-1,N)$. More specifically, this result implies that $\mathrm{rank}\bigl(\ma{\Theta}\bigr)=N$ if $\mathrm{rank}(\mat{G})+\mathrm{rank}(\mat{H}) \geq N+1$, which is ensured if $\mat{G}^\trans$ or $\mat{H}$ have full column-rank. According to this, the uniqueness of $\hat{\mat{X}}$ implies $KM \geq L$. The practical impact of this result is that the symbol estimation is still ensured in more difficult scenarios with poor scattering for the RIS-BS link (e.g. millimeter wave or Terahertz frequencies), where $\mat{H}$ is likely rank-deficient, i.e., $(1\leq \mathrm{rank}\bigl(\mat{H}\bigr) < \mathrm{min}(M,N))$. In the worst-case scenario, where $\mat{G}$ and/or $\mat{H}$ are rank-deficient, the rank of each $k$-th block of $\mat{E}(\ma{\Theta})$ is equal to 1. As a result, the UT must transmit at least $K\geq L$ sub-frames to ensure that $\mat{E}(\ma{\Theta})$ has full column-rank\footnote{According to \cite{favier2020algebraic}, $\mathrm{rank}\bigl(\mat{E}(\ma{\Theta})\bigr) \!=\!\textstyle \sum_{k=1}^K{\!\mathrm{rank}(\mat{E}_k)}$ under the constraint $\mathrm{dim}[\mathtt{R}(\mat{E}_1) \cap \cdots \cap \mathtt{R}(\mat{E}_K)]=0$, where $\mathtt{R}(\mat{E}_k)$ denotes the row-space of $\mat{E}_k$. This constraint can be achieved by properly designing $\vet{\psi}_k$ and $\vet{\lambda}_k$ ($k\!=\!1,\!\cdots\!,K$). For $\mat{E}(\ma{\Theta})$ to have full column-rank, the sum of the ranks of all blocks $\mat{E}_1, \ldots, \mat{E}_K$ must be at least $L$. Therefore, if $\mathrm{rank}(\mat{E}_k)\!=\!1$, then, $\mathrm{rank}\bigl(\mat{E}(\ma{\Theta})\bigr) \!=\! \mathrm{min}\{\textstyle \sum_{k=1}^K{\!\mathrm{rank}(\mat{E}_k)},L\} = \mathrm{min}\{K,L\} = L$, if $K \geq L$.}.

In summary, satisfying the conditions $KT \!\geq\! NL$ and $KM \!\geq\! L$ is equivalent to $K \!\geq\! L\!\lowint{\mathrm{max}\{N/T,1/M\}}$, which establishes the minimum number of sub-frames to ensure joint estimation of the combined channel and transmitted symbols.

\subsection{Scaling ambiguities}
Under the conditions above, the estimates of $\ma{\Theta}$ and $\mat{X}$ after convergence of the BALS are unique up to scaling ambiguities, i.e., $\hat{\mat{X}} \!=\! \ma{\Delta}_\mathrm{x}\mat{X}$ and $\hat{\ma{\Theta}} \!=\! \ma{\Delta}_\theta\ma{\Theta}$, where $\ma{\Delta}_\mathrm{x} \!\in\! \compl^{L \times L}$ and $\ma{\Delta}_\theta \!\in\! \compl^{LM \times LM}$ are diagonal matrices that cancel each other. By inserting $\ma{\Delta}_\mathrm{x}$ and $\ma{\Delta}_\theta$ in \eqref{ykthetapsik}, we get $\vet{y}_k =$ $\bigl(\mat{X}^\trans\!\ma{\Delta}_\mathrm{x}\diagof{\vet{\lambda}_k} \!\kron\! \mat{I}_M\bigr)\ma{\Delta}_\theta\ma{\Theta}\vet{\psi}_k \!+\! \vet{v}_k$. Interchanging the positions of $\ma{\Delta}_\mathrm{x}$ and $\diagof{\vet{\lambda}_k}$ and using the property \eqref{prop:kronkron}, yields
\begin{equation}
    \vet{y}_k = \bigl(\mat{X}^\trans\diagof{\vet{\lambda}_k} \kron \mat{I}_M\bigr)\bigl(\ma{\Delta}_\mathrm{x} \kron \mat{I}_M\bigr)\ma{\Delta}_\theta\ma{\Theta}\vet{\psi}_k + \vet{v}_k.
\end{equation}
Therefore, $\ma{\Delta}_\theta = (\ma{\Delta}_\mathrm{x}^{-1} \kron \mat{I}_M)$. To resolve this ambiguity, a pilot symbol should be transmitted at each antenna at the first symbol period, implying the first column of $\mat{X}$ be known, e.g., $\vet{x}_1 \!=\! [1,\!\cdots\!,1]^\trans \!\in\! \compl^{L \times 1}$. Note that the estimates of $\mat{G}$ and $\mat{H}$ delivered by the KRF stage are affected by diagonal scaling matrices, satisfying $\hat{\mat{G}} \!=\! \ma{\Delta}_\mathrm{g}\mat{G}$ and $\hat{\mat{H}} \!=\! \mat{H}\ma{\Delta}_\mathrm{h}$, so that $\ma{\Delta}_\mathrm{g}\ma{\Delta}_\mathrm{h} = \mat{I}_N$. However, such ambiguities are irrelevant since they compensate each other while system optimization is usually carried out from the combined channel \cite{zappone2021overhead}.

\section{Simulation results}
We evaluate the performance of the proposed TSB receiver in terms of the normalized mean square error (NMSE) of the channels, given by $\textrm{NMSE}(\hat{\ma{\Theta}}) = \|\ma{\Theta} - \hat{\ma{\Theta}}\|_\mathrm{F}^2/\|\ma{\Theta}\|_\mathrm{F}^2$, the symbol error rate (SER), and the computational complexity. The symbols follow a 64-QAM constellation while the channel matrices are modeled according to the Saleh-Valenzuela model \cite{zhou2022channel}, with the number of paths set to 1 for UT-RIS and RIS-BS links, capturing the characteristics of mmWave propagation. We jointly design the coding and phase-shift vectors from a truncated DFT matrix according to \cite{sokal2020semi,dearaujo2023semiblind}. We consider the setup $\{M,N,L,T,K\} = \{8,32,2,4,64\}$ and the results are averaged over $R=10^4$ Monte Carlo runs.

\begin{figure}[!t]
	\centering
	\includegraphics[width=0.4\textwidth]{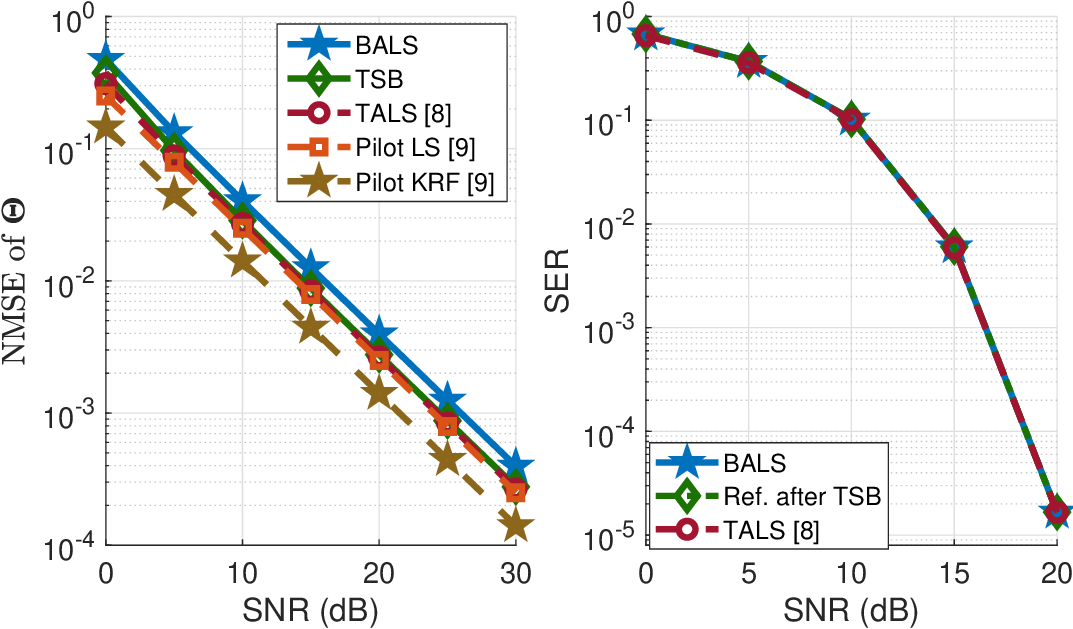}
    \vspace{-1ex}
    \caption{NMSE of the combined channel and symbol error rate vs. SNR (dB).}
	\label{fig:nmseser}
\end{figure}

Figure \ref{fig:nmseser} shows the NMSE results for the combined channel and the SER as a function of the SNR. From the CE perspective, the gap between the BALS stage and the competitor of \cite{dearaujo2023semiblind} is only 1.8 dB. Note, however, that applying KRF to extract separate estimates of $\hat{\mat{G}}$ and $\hat{\mat{H}}$ followed by reconstructing $\hat{\ma{\Theta}}$ enhances the estimation accuracy, making TSB perform similar to TALS. This improvement comes from the increased noise rejection when decoupling the channel estimates, corroborating the analysis in \cite{dearaujo2021channel}. We consider the pilot-aided LS and the KRF methods detailed in \cite{dearaujo2021channel} as a comparison reference. Figure \ref{fig:nmseser} shows similar SER performances for all methods. This is expected since symbol estimation has already experienced a refinement at each iteration for both iterative methods. Therefore, the KRF stage is important for refining $\hat{\ma{\Theta}}$ and is unnecessary for $\hat{\mat{X}}$, as we can see in the result ``Ref. after TSB'', after using $\hat{\mat{G}}$ and $\hat{\mat{H}}$ to refine $\hat{\mat{X}}$.

\begin{figure}[!t]
	\centering
	\includegraphics[width=0.4\textwidth]{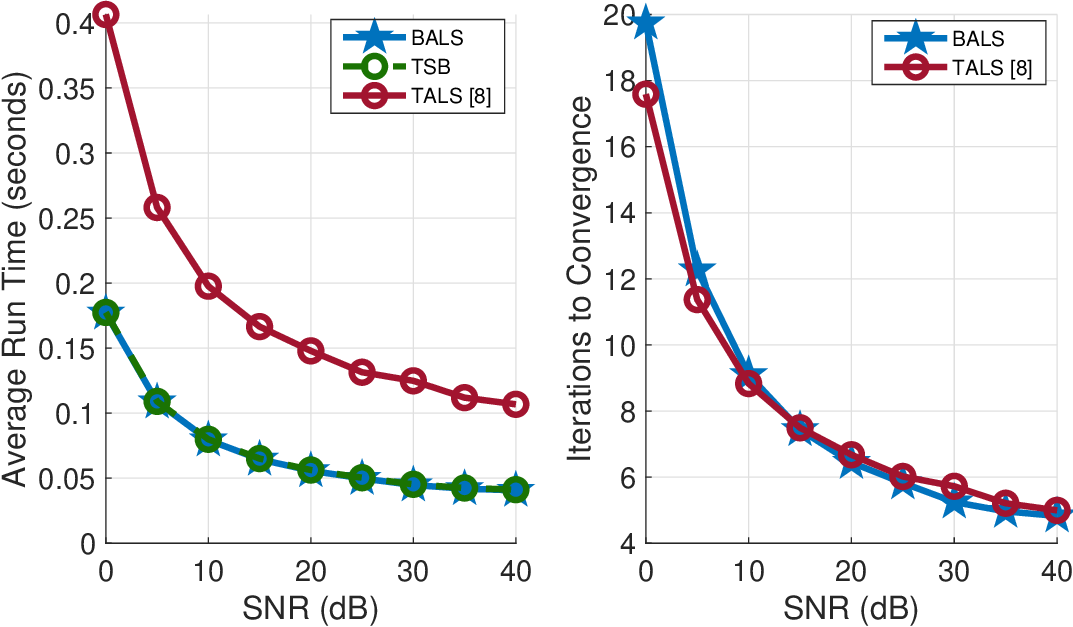}
    \vspace{-1ex}
    \caption{Average run time and iterations to convergence vs. SNR (dB).}
	\label{fig:timeit}
\end{figure}

\begin{figure}[!t]
	\centering
    \includegraphics[width=0.34\textwidth]{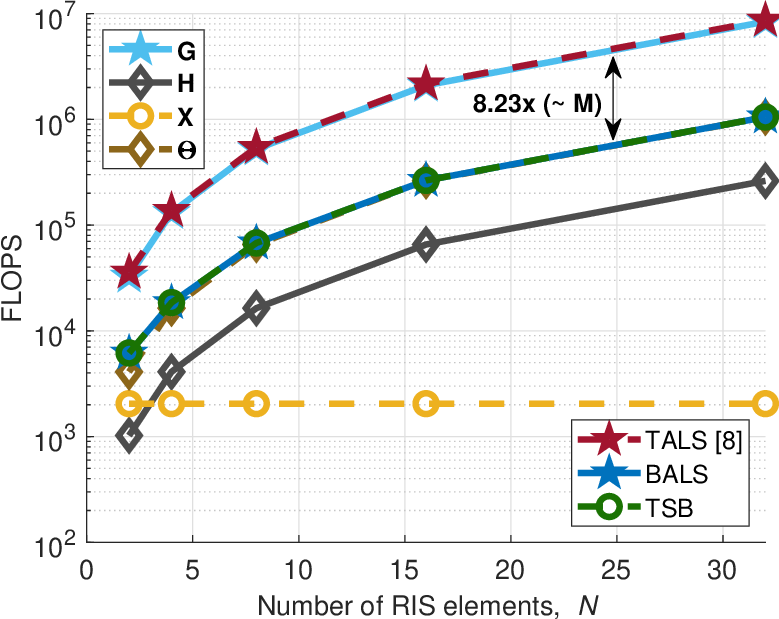}
     \vspace{-1ex}
   \caption{Number of FLOPS vs. number of RIS elements.}
	\label{fig:flops}
\end{figure}

Figure \ref{fig:timeit} shows the results of the average run time and number of iterations to convergence. We can see that the convergence of BALS (first stage of the proposed TSB receiver) is practically identical to that of the TALS competitor \cite{dearaujo2023semiblind}, even in the lower SNR range (<\! 10 dB). This implies, consequently, a lower run time to BALS. To clarify, three LS problems are solved in TALS \cite{dearaujo2023semiblind} to find $\hat{\mat{G}}$ and $\hat{\mat{H}}$ (to construct $\hat{\ma{\Theta}}$), and $\hat{\mat{X}}$. Conversely, the first stage of the proposed TSB receiver requires solving only two LS problems to get $\hat{\ma{\Theta}}$ and $\hat{\mat{X}}$. A negligible increase in the run time was observed after running the second stage KRF (see the ``TSB'' curve). This leaves room to compare the computational complexity of both methods by assuming the same number of iterations for both. A direct way to make this study is to measure the complexity of a single iteration as a reference. To this end, in Figure \ref{fig:flops}, we evaluate the complexity of the TSB receiver as compared to that of TALS in terms of floating point operations (FLOPS). We also plot the complexities of estimating the individual matrices as a reference. The results show that the TSB receiver requires significantly fewer operations than its competitor. Indeed, estimating $\mat{H}$ and $\mat{G}$ separately in an alternating way (which is the case with the competitor TALS \cite{dearaujo2023semiblind}) leads to costs of $\mathcal{O}(KTN^2)$ and $\mathcal{O}(MKTL^2N^2)$ per iteration, respectively. In particular, note that the cost of estimating the individual UT-RIS channel matrix $\mat{G}$ by itself in the TALS competitor already surpasses by $M$ times the overall complexity involved in getting an estimate of $\ma{\Theta}$ in the proposed TSB receiver. As seen in Figure \ref{fig:flops}, a complexity reduction of 8.2 times is achieved. These results clearly show that estimating the combined channel semi-blindly, followed by channel decoupling, is an effective solution to RIS-assisted MIMO systems since it provides a significant complexity reduction with negligible performance loss compared to the competing baseline data-aided method, as can be concluded from the results shown in this section. It is worth noting that the algebraic steps to derive the TSB receiver can be adapted to coding structures differing from the Khatri-Rao coding scheme, such as a more general tensor ST coding scheme \cite{favier2012tensor}. This generalization will be considered in a future work.


\section{Conclusion}\label{sec:conclusion}
We have derived a novel tensor-based data-aided semi-blind (TSB) receiver for RIS-assisted MIMO communications capable of decoupling the estimation of the composite channel from the data symbols. The proposed receiver is formulated by recasting an iterative PARATUCK-2-based ALS algorithm as a simpler two-stage estimation approach. Our results show that the TSB receiver has significantly less computational cost than the competing data-aided receiver while offering similar channel estimation accuracy using actual symbol estimates. Therefore, the proposed TSB receiver is a competing solution to RIS-assisted data-aided channel estimation.

\renewcommand\baselinestretch{.91}
\bibliographystyle{IEEEtran}
\bibliography{references}
\end{document}